\documentclass[usenatbib]{mn2e}

\usepackage{graphicx}
\usepackage{amsmath}
\usepackage{amssymb}
\usepackage{url}
\usepackage{multirow}

\usepackage[norounding]{rccol}
\rcDecimalSign{.}

\newcommand{\nan}{\multicolumn{1}{c}{---}}

\newcommand{\flux}   {erg cm$^{-2}$ s$^{-1}$}

\def\gsim{ \lower .75ex \hbox{$\sim$} \llap{\raise .27ex \hbox{$>$}} }
\def\lsim{ \lower .75ex\hbox{$\sim$} \llap{\raise .27ex \hbox{$<$}} }

\def\sc{Schwarzschild}

\def\srcname{B2 0954+25A}

\title[\srcname: a typical blazar or a $\gamma$--NLS1?]  
{\srcname: a typical {\it Fermi} blazar or a $\gamma$--ray loud Narrow
  Line Seyfert 1?}

\author[G. Calderone et al.]
  {G. Calderone$^{1}$\thanks{E-mail: {\tt giorgio.calderone@mib.infn.it}},
  G. Ghisellini$^{2}$, 
  M. Colpi$^{1}$,
  M. Dotti$^{1}$\\
  $^{1}$Universit\`a di Milano - Bicocca, Dip. di Fisica G. Occhialini, Piazza della Scienza 3, I-20126 Milano, Italy\\
  $^{2}$INAF Osservatorio Astronomico di Brera, Via E. Bianchi 46, I-23807 Merate (LC), Italy\\ 
}

\begin{document}

\pagerange{\pageref{firstpage}--\pageref{lastpage}} \pubyear{2011}

\maketitle

\label{firstpage}

\begin{abstract}
\srcname, detected by the {\it Fermi} satellite, is a blazar with
interesting observational properties: it has been observed to transit
from a jet dominated to a disk dominated state; its radio spectrum
appears flat at all observing frequencies (down to 74 MHz); optically,
the H$\beta$ line profile is asymmetric. The flatness of radio
spectrum suggests that the isotropic emission from radio lobes is very
weak, despite the large size of its jet ($\gtrsim$ 500 kpc).  Its
broad--band spectral energy distribution is surprisingly similar to
that of the prototypical $\gamma$--ray, radio loud, Narrow Line
Seyfert 1 ($\gamma$--NLS1) galaxy PMN J0948+0022.  In this work we
revisit the mass estimates of \srcname\ considering only the symmetric
component of the H$\beta$ line and find (1--3) $\times 10^8$
M$_{\sun}$.  In light of our composite analysis, we propose to
classify the source as a transition object between the class of Flat
Spectrum Radio Quasar and $\gamma$--ray, radio loud NLS1. A comparison
with two members of each class (3C 273 and PMN J0948+0022) is
discussed.
\end{abstract}


\section{Introduction}

A major breakthrough in the comprehension of the AGN phenomenon has
been the formalization of the so--called ``unified model''
\citep{1995-Urry-unifiedscheme}.  All active galactic nuclei are now
believed to be similar objects, whose gross observational features
depend on the black hole mass, the accretion rate, the viewing angle
and the presence of a jet \citep{1994-Rawlings-trulyunified}.  Our
understanding of finer details is, however, far from complete: each
source has its own list of peculiarities that are difficult to explain
coherently in the framework of the unified model.  Furthermore, the
values of the black hole mass, that can be inferred only indirectly,
are still uncertain.  Identification and analysis of peculiar features
in single sources is thus important to improve our understanding of
the AGN phenomenon.  In this context the {\it Fermi} blazar
\srcname\ (= OK 290 = TXS 0953+254), detected in the $\gamma$--ray
band by the {\it Fermi} satellite, has several interesting features
worth investigating.  Interest for this source has been stimulated by
the wide range of values for its black hole mass and width of hydrogen
Balmer lines found in literature.

At optical wavelengths the source has been observed in at least three
emission states: the partially jet dominated state in 1987
\citep{1991-jackson-quasarprop}, as observed by the {\it Isaac Newton
  Telescope} (INT); the disk dominated state in 2004, as observed by
the Sloan Digital Sky Survey \citep[SDSS, ][]{2009-sdss-dr7}; and the
completely jet dominated state in 2006, observed again with the SDSS.
The optical spectrum reveals that the H$\beta$ emission line shows an
asymmetry in its red wing, as observed in many AGNs.  This asymmetry
is one of the likely reasons for the different values of the black
hole mass found in the literature.  Depending on the decomposition
made to analyze the optical spectrum, the FWHM of the H$\beta$ broad
line may be lower than the threshold of 2000 km s$^{-1}$, commonly
used to classify Narrow Line Seyfert 1 (NLS1) sources, and the
resulting black hole mass would be $\sim10^8 M_{\sun}$ (\S
\ref{sec-spec-derived-res}).  Although there is nothing special in the
2000 km s$^{-1}$ threshold
\citep[e.g. ][]{1989-Goodrich-spectropolarimetry-NLS1,
  2001-Veron-AtlasOfNLS1}, we believe that such a low mass provides a
hint for a resemblance of \srcname with NLS1 sources (\S
\ref{sec-sed-compare}).  Indeed, its spectral energy distribution
(SED) is almost identical to PMN J0948+0022
\citep{2009-Abdo-mw_monitor_pmnj0948,
  2009-Foschini-discovery_pmnj0948}, the first radio loud NLS1
detected by {\it Fermi}.  In this case the source would join the small
group of radio--loud NLS1 that have been detected by the {\it Fermi}
satellite \citep{2009-Abdo-discovery_pmnj0948,
  2009-Abdo-rlnls1_newclass, 2011-foschini-procnls1}.

In this work we present a multi--wavelength study on \srcname, with
particular emphasis on the spectral analysis at optical wavelengths,
broad--band SED modeling and radio properties.  Our aim is to provide
a coherent picture of its physical properties, according to
observational data. Our analysis relies on data from several
facilities, as discussed in the following sections. General data are
shown in Tab. \ref{tab-datapos}.
\begin{table}
  \begin{minipage}{80cm}
    \caption{General data.}
    \label{tab-datapos}
    \begin{tabular}{lR{6}{3}c}
      \hline\hline
      \multicolumn{1}{c}{Parameter}          &
      \multicolumn{1}{c}{Value}              &
      \multicolumn{1}{c}{Units}              \\
      \hline
R.A.                                                                                 &149.208 &deg               \\
Declination                                                                          &25.254  &deg               \\
$z$\footnote{From [OIII] line fitting (see \S \ref{sec-optspectro})}                 &0.70747 &                  \\
$d_{\rm L}$                                                                          &4.3     &Gpc               \\
$A_{\rm B}$\footnote{Galactic absorption from \citet{1998-Schlegel-galaxy-irmap}}    &0.158   &                  \\
E(B--V)$^b$                                                                          &0.0375  &                  \\
$N_{\rm H}$\footnote{Computed using the LAB survey \citep{2005-kalberla-labsurbey}}  &3.      &$10^{20}$ cm$^{-2}$\\
      \hline\hline      
    \end{tabular}
  \end{minipage}
\end{table}


Throughout the paper, we assume a $\Lambda$CDM cosmology with H$_0$ =
71 km s$^{-1}$ Mpc$^{-1}$, $\Omega_{\rm m}$ = 0.27, $\Omega_\Lambda$ =
0.73.

\section{The source \srcname}
\label{sec-intro-b2}

The source \srcname\ ($z$=0.712,
\citealt{1972-Burbidge-SpectroscopyOf22QSO}) is a compact,
radio--loud, flat--spectrum radio quasar (FSRQ).  It has been
frequently used in statistical studies on radio properties of quasars
since it is a relatively luminous radio source \citep[$\gtrsim$1 Jy,
][]{1981-kuhn-sample1jy}, whose emission extends to very low
radio--frequencies (74 MHz, \citealt{2007-VLA-survey}).  The radio
spectrum is usually flat and becomes inverted during burst activity
\citep{2005-Torniainen-VariabilityOfGPS}.  The jet is clearly visible
in several radio maps: see e.g. the VLA radio maps at 1.64 GHz in
\citet{1993-murphy-vlacompletesample} and the VLBA radio maps at 22
and 43 GHz in \citet{2000-lister-pc-vs-kpc-scale-jets}. The core
component has angular size $\sim0.23 \times 0.07$ mas (at 15 GHz),
corresponding to a linear size of $1.6 \times 0.49$ pc
\citep{2005-Kovalev-SubMasImagingOfQuasars}. Several components in the
jet show superluminal motion (up to 12$c$,
\citealt{2004-Kellermann-SubMasImagingOfQSO}).  A one--sided jet
(projected size $\sim$50 kpc) extends from the core in the south--west
direction \citep{2002-Liu-jetlist}.

In the optical band, the source is unresolved, variable
\citep{1988-Pica-VariabilityOfCompactQSO} and slightly polarized
\citep[1.29\%,][]{1992-Wills-PolarizationSurvey}, with $m_i \sim$18
mag.  The bolometric luminosity (estimated from SED fitting,
\citealt{2002-Woo-BHMassAndLum}) is log($L_{\rm bol}$/erg s$^{-1}$) =
46.59.  Virial black hole mass estimates are log($M/M_{\sun}$)=8.7
\citep{2006-Liu-JetPowerRL-BHM} and log($M/M_{\sun}$)=9.5
\citep{2001-Gu-BHMassesInRQ}.  Both these estimates rely on a FWHM
estimate of the H$\beta$ emission line given in
\citet{1991-jackson-quasarprop}, who found FWHM(H$\beta$) =
65 \AA\ (rest frame), corresponding to $\sim$4000 km s$^{-1}$.  The
source is also present in the \citet[][hereafter
  S10]{2010-shen-catdr7} catalog, which contains several measures
obtained by automatically analyzing the SDSS/DR7 spectrum.  S10 report
a FWHM(H$\beta$) = 1870 km s$^{-1}$ and log($M/M_{\sun}$)=8.6
(computed with H$\beta$), or 9.3 (computed with MgII).  The spectrum
analyzed in S10 is exactly the same as the one we use in \S
\ref{sec-optspectro}.

Several X--ray facilities ({\it Swift}/XRT, {\it Chandra}, {\it
ROSAT}, {\it Einstein}) observed the source at different times
measuring fluxes in the range (2.5--20) $\times 10^{-13}$ erg
cm$^{-2}$ s$^{-1}$ (\S \ref{sub-sed-data} and Fig. \ref{fig-sed}).
Finally, \srcname\ is present in both the 1yr and 2yr {\it Fermi}
Large Area Telescope (LAT) point source catalogs \citep[][with catalog
names 1FGL J0956.9+2513 and 2FGL J0956.9+2516
respectively]{2010-1FGL, 2011-lat2fgl}.

\subsection{Archival data}
\label{sub-sed-data}

To build the broad--band SED (Fig. \ref{fig-sed}) we collected
publicly available data from several facilities using
NED.\footnote{Nasa/IPAC Extragalactic Database
  \url{http://http://ned.ipac.caltech.edu/}} Tab. \ref{tab-dateobs}
shows the observation dates for all facilities.  Here we provide a
brief description of data analysis for each of the facilities:
\begin{figure*}
\vskip -0.5 cm
\includegraphics[width=17cm]{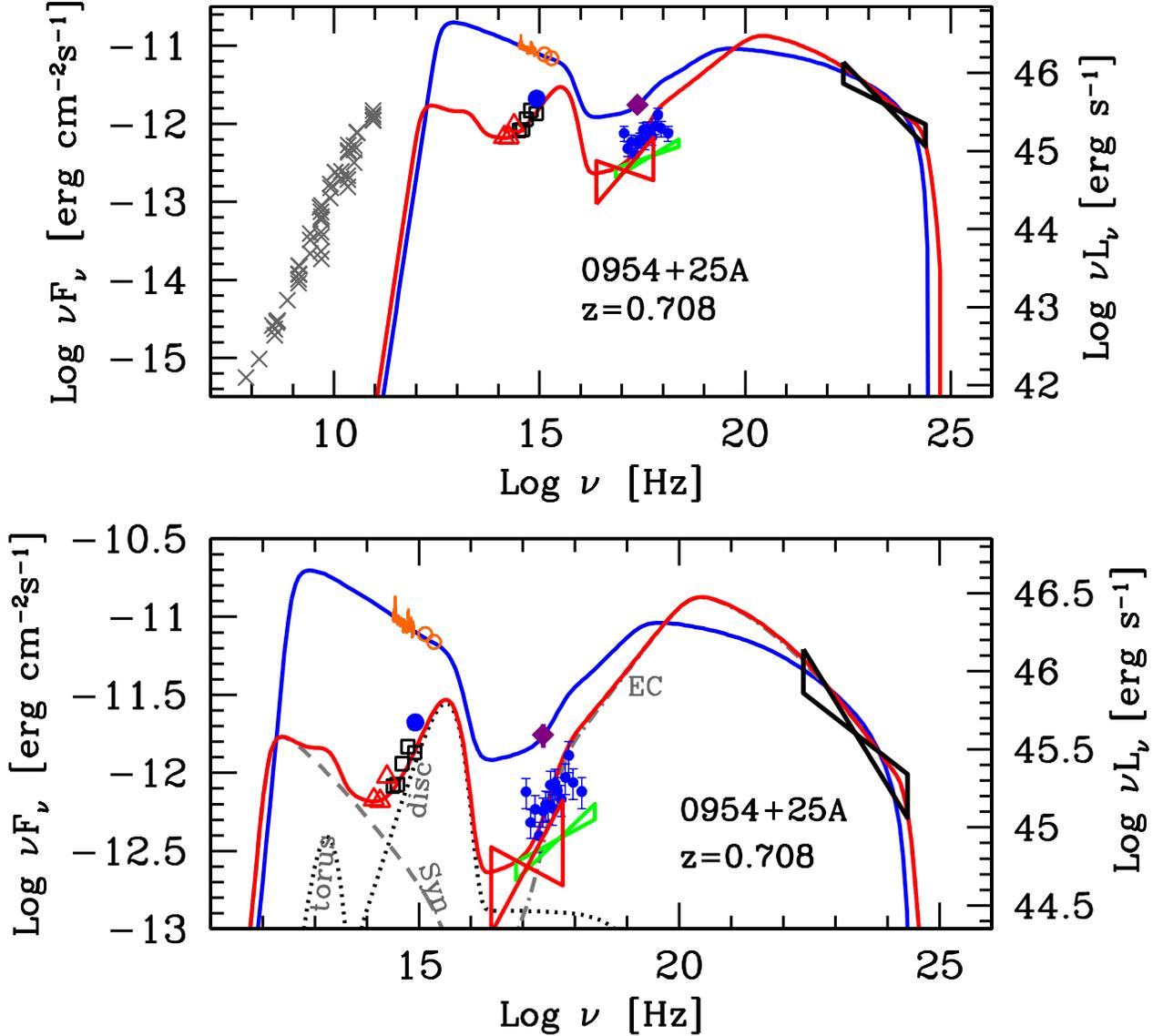}
\vskip -0.7 cm 
\caption{Broad band spectral energy distribution, and related models,
  of \srcname. Data are provided by several facilities (see
  \S\ref{sub-sed-data}). Upper panel shows all available data, from
  radio to $\gamma$--ray energies, as well as the SED models in the
  low and high synchrotron state. The lower panel shows a detailed
  view of the SED from IR to $\gamma$--rays, and highlights the most
  important components of the models.}
\label{fig-sed} 
\end{figure*}

\begin{table}
  \center
  \caption{Observation dates for all data used to build the SED in
    Fig. \ref{fig-sed}.}
   \label{tab-dateobs}
   \begin{tabular}{lc}
     \hline\hline
     Facility              & Date of obs.\\
     \hline
     {\it Einstein}        & 1979--11--06\\
     INT                   & 1987--12--16\\
     2MASS                 & 1998--11--30\\
     SDSS (photometry)     & 2004--12--13\\
     SDSS (spectrum)       & 2006--01--05\\
     {\it GALEX}           & 2006--02--05\\
     {\it Swift}           & 2007--06--01\\
     {\it Chandra}         & 2009--01--20\\
     {\it Fermi} (average) & Aug 2008 -- Jun 2011 \\
     \hline\hline
   \end{tabular}
\end{table}

\vskip 0.3 cm
\noindent 
{\it Fermi/LAT ---}
\label{sec-data-fermi}
The 2LAC catalog \citep{2011-ackermann-2lac} reports a flux for the
source of $(2.1 \pm 0.2) \times 10^{-12}$ erg s$^{-1}$ cm$^{-2}$ at
618 MeV, and a photon index $\Gamma = -2.39 \pm 0.08$.  We also ran a
likelihood analysis on the first 34 months of data collected by {\it
  Fermi}/LAT (using {\it ScienceTools} ver. 9.23.1 and Instrument
Response Function P6\_V3) with time bins of one month, to check for
the occurrence of major flares. We found no significant variability of
the flux nor of the spectral index, i.e. a constant flux and spectral
index model fits well observational data and cannot be rejected on
statistical basis.  The flux in the 0.1--100 GeV band averaged over 34
months is compatible with the value reported in 2LAC.

\vskip 0.3 cm
\noindent 
{\it Swift/XRT ---} On June 1st, 2007 {\it Swift} performed a pointed
observation (obsID 00036325002) of a nearby source, thus it has been
possible to extract data for \srcname.  We extracted a spectrum using
the \verb|xrtpipeline| script and binned it to have at least 20 counts
in each bin.  Then we used XSPEC to fit a simple power law with the
hydrogen column fixed to the Galactic value $N_{\rm H}=3.2\times
10^{20}$ cm$^{-2}$.  The resulting de--absorbed flux in the 0.3--10
keV energy band is $(2.5\pm 0.3)\times 10^{-12}$ \flux and the photon
index is $\Gamma = 1.74 \pm 0.09$.

\vskip 0.3 cm
\noindent
{\it Chandra ---}
On June 20th, 2009 {\it Chandra} pointed our source.  We extracted a
spectrum using the \verb|specextract| script, with a minimum of 30
counts in each bin, then performed a fit against a simple power law
using {\it Sherpa}.  Again the $N_{\rm H}$ column has been kept fixed
to $3.2\times 10^{20}$ cm$^{-2}$.  The resulting de--absorbed flux in
the 0.3--10 keV energy band is $(1.32\pm 0.06)\times 10^{-12}$ \flux
and the photon index is $\Gamma = 1.75 \pm 0.07$.

\vskip 0.3 cm
\noindent
{\it GALEX ---}
on Feb 5th, 2006 {\it GALEX} observed the source and provided the
following photometric measurements: $(269 \pm 18)$ $\mu$Jy at 1528
\AA\ (far UV) and $(535 \pm 14)$ $\mu$Jy at 2271 \AA\ (near UV).
Using a colour excess E(B--V) = 0.0377 we computed the de--reddened
fluxes using CCM 1989 parametrization
\citep{1989-cardelli-extinction}.  The resulting $\nu F_\nu$ values
are $(7.00 \pm 0.47)\times 10^{-12}$ \flux (FUV) and $(9.71 \pm
0.26)\times 10^{-12}$ \flux (NUV).

\vskip 0.3 cm
\noindent
{\it Swift/UVOT ---}
An aperture photometry analysis of the $u$ filter data provided a
count rate of 3.16 $\pm 0.11$ s$^{-1}$.  The corresponding
\citep{2008-poole-uvotcalib} de--reddened
\citep{1989-cardelli-extinction} $\nu F_\nu$ flux at 3501 \AA\ is
$(2.14\pm 0.07)\times 10^{-12}$ \flux.

\vskip 0.3 cm
\noindent
{\it SDSS ---} Our source has been observed photometrically
(Dec. 2004) and spectroscopically (Jan. 2006, see also \S
\ref{sec-optspectro}) by SDSS.  Both data sets have been de-reddened
using \citet{1989-cardelli-extinction}.

\vskip 0.3 cm
\noindent
{\it 2MASS ---}
An IR observation of the source has been performed on Nov 30th, 1998
and results are available in the 2MASS catalog.  Magnitudes are $m_J
=16.53 \pm 0.13$, $m_H = 16.12 \pm 0.19$ and $m_K = 15.34 \pm 0.14$.
The corresponding
\footnote{\url{http://www.ipac.caltech.edu/2mass/releases/allsky/faq.html#jansky}}
$\nu F_\nu$ fluxes are: $(9.5\pm 1.1)\times 10^{-13}$ \flux at a 1.2
$\mu$m (J band), $(6.6\pm 1.1)\times 10^{-13}$ \flux at a 1.6 $\mu$m
(H band) and $(6.8\pm 0.8)\times 10^{-13}$ \flux at a 2.2 $\mu$m (K
band).

\bigskip

The broad band SED (built using archival data from facilities
described in \S \ref{sub-sed-data}) is shown in Fig. \ref{fig-sed}.
Grey crosses are from radio facilities, black squares are from Sloan
photometry, red butterfly is from {\it ROSAT}, purple diamond is from
{\it Einstein}.  We added the data from {\it Fermi}/Large Area
Telescope (LAT, black butterflies), {\it Swift}/X--Ray Telescope (XRT,
blue dots), {\it Chandra} (green butterfly), {\it GALEX} (orange open
circles), {\it Swift}/UltraViolet Optical Telescope (UVOT) (blue
circles) and 2MASS (red triangles).  Colors of simultaneous
observations ({\it Swift}/XRT and UVOT, SDSS spectroscopic and {\it
  GALEX}) are the same.  Looking at the SED in optical/UV it is clear
that the source raised its flux by a factor $\sim$10 in about one
year.  In June 2007 the source returned to a lower state (in
optical/UV) as shown by the {\it Swift}/UVOT measurements.  Thus, the
source shows at least a ``low" and a ``high'' state in optical/UV.  A
similar trend, although of smaller magnitude, is shown at X--ray
wavelengths with the lowest flux being measured by {\it ROSAT} and the
highest being measured by {\it Einstein}.

In the following sections we will focus on specific wavelengths
range, namely radio (\S \ref{sec-radioprop-0954}) and optical (\S
\ref{sec-optspectro}), and then provide an overall picture of the
broad band SED (\S \ref{sec-sed-modeling}).

\section{Radio properties}
\label{sec-radioprop-0954}

\srcname\ shows a flat radio spectrum from $\sim 10^8$ Hz to $\sim
10^{11}$ Hz.  At 5 GHz, the luminosity of the core is log($L_{\rm
  c}$/erg s$^{-1}$ Hz$^{-1}$) = 34.5, while that of the extended
region is log($L_{\rm e}$/erg s$^{-1}$ Hz$^{-1}$) = 32.4, thus the
source is highly core--dominated with a core dominance parameter
$R_{\rm cd} \sim 100$, i.e. at the higher end of the core--dominance
parameter distribution
\citep{1987-Browne-BeamingXrayOpticalAndRadioPropOfQSO}.  The weakness
of the extended emission is also evident at low radio frequencies,
because the radio spectrum remains flat down to 74 MHz, with a
spectral index $\alpha \sim -0.2$ between 74 and 365 MHz ($F_\nu
\propto \nu^{\alpha}$).  Assuming a spectral index for the optically
thin emission in the extended region of --0.7, and extrapolating the
flux from 5 GHz, the frequency at which the isotropic and the beamed
jet luminosities are equal is $\sim$10 MHz, in agreement with the
observed overall flat radio spectral index.  This suggests that the
jet is well aligned along the line of sight.

\begin{figure}
\hskip -0.7 cm
\includegraphics[width=9.7cm]{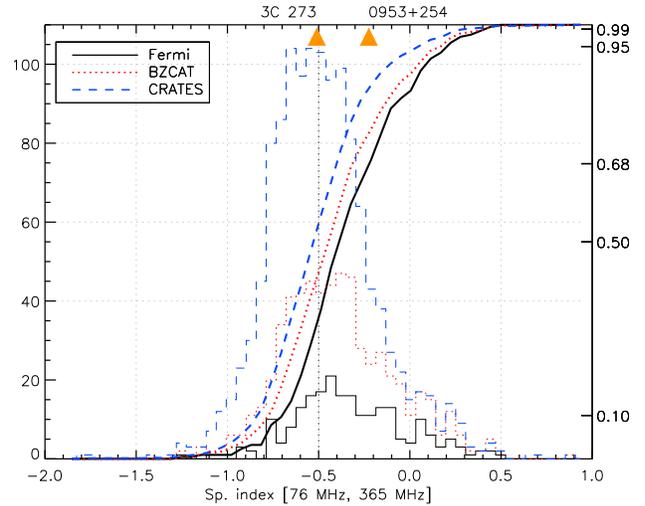} 
\caption{Distribution of spectral index between the (observed)
  frequencies 74 and 365 MHz for three different blazar catalogs:
  CRATES \citep[][blue dashed line]{2007-Healey-CRATES}, BZCAT
  \citep[][red dotted line]{2009-Massaro-BZCAT} and {\it Fermi} 2yr
  catalog of AGN \citep[][2LAC, black solid
    line]{2011-ackermann-2lac}. Thicker lines show cumulative fraction
  (values on right axis). Radio fluxes to compute the spectral index
  are taken from the VLA survey at 74 MHz \citep{2007-VLA-survey} and
  the Texas survey at 365 MHz \citep{1996-Douglas-TEXAS}.}
\label{fig-radiospindex}
\end{figure}

We can compare the 74--365 MHz radio spectral index $\alpha$ of
\srcname\ with the distribution of the same spectral index of other
blazars.  To this aim we show in Fig. \ref{fig-radiospindex} the
distribution of spectral index between the (observed) frequencies 74
and 365 MHz for three different blazar catalogs: CRATES \citep[][blue
  line]{2007-Healey-CRATES}, BZCAT \citep[][red
  line]{2009-Massaro-BZCAT} and {\it Fermi} 2yr source catalog of AGN
\citep[2LAC, all ``CLEAN'' sources, ][black
  line]{2011-ackermann-2lac}.  The radio fluxes to compute the
spectral index are taken from the VLA survey at 74 MHz
\citep{2007-VLA-survey} and the Texas survey at 365 MHz
\citep{1996-Douglas-TEXAS}. The number of sources detected at both
frequencies is 1282 for the CRATES catalog; 605 for the BZCAT and 218
for the {\it Fermi} 2LAC catalog.  The distribution shows that only
$\sim$30\% of the highly beamed blazars detected by {\it Fermi} have a
steep ($\alpha<-0.5$) spectral index between 74 and 365 MHz.
Radio--selected blazars from CRATES show a broader distribution with
$\sim$50\% sources having spectral index $<$--0.5.  The BZCAT
distribution is similar to CRATES, although with a larger fraction of
very flat or inverted radio sources.

The extended region of \srcname, as observed in the 1.64 GHz VLA radio
map, has a projected size of $\sim$50 kpc
\citep{1993-murphy-vlacompletesample, 2002-Liu-jetlist}.  This is
rather large, especially considering that the viewing angle
$\theta_{\rm v}$ is small, as suggested by the observed superluminal
motion \citep[12$c$, ][]{2004-Kellermann-SubMasImagingOfQSO} and by
the SED modeling (\S 3.5 and Tab. \ref{para}).  Assuming $\theta_{\rm
  v}$ in the range 3--6 degrees, the de--projected size is then in the
range 0.5--1 Mpc, typically the size of a giant radio lobe.

We conclude that the radio properties of \srcname\ indicate a highly
core dominated source, with an extended component that, albeit
reaching the dimension of a giant radio lobe, is rather weak, making
the core dominance parameter very large, at the high end of the
distribution values for blazars.

\section{Optical spectroscopy}
\label{sec-optspectro}

\begin{figure*}
\includegraphics[width=12.8cm]{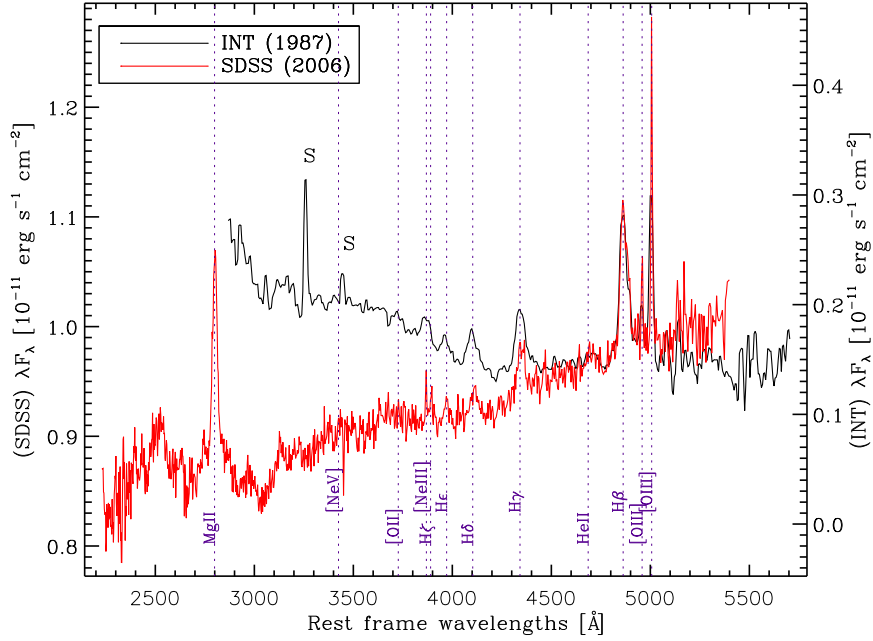}
\caption{Optical spectrum of \srcname\ observed by SDSS in 2006 (red
  line, flux scale on the left axis) and with the {\it Isaac Newton
    Telescope} (INT) in 1987 (black line, flux scale on the right
  axis).  The range of values covered by both axes is exactly the same
  ($\sim 0.5 \times 10^{-11}$ erg s$^{-1}$ cm$^{-2}$) so that spectra
  can be directly compared.  SDSS spectrum have been de--reddened
  using \citet{1989-cardelli-extinction} with E(B--V)=0.0375,
  transformed to the rest frame using our redshift estimate
  ($z$=0.70747, see text) and rebinned by a factor of 2.  The INT
  spectrum has been derived from the plot given in \citet[][their
    Fig. 2]{1991-jackson-quasarprop} and transformed to rest frame.
  The ``S" letter denote a telluric line.  Some of the most important
  emission lines are highlighted.  The spectra show that the continuum
  has changed both in intensity (by a factor $\sim$5) and in slope.
  The intensity and shape of the Balmer lines seem to be the same in
  both spectra.}
\label{fig-spec-werr} 
\end{figure*}

We used optical spectra from the {\it Isaac Newton Telescope} (INT)
and from SDSS\footnote{http://www.sdss.org/dr7. Data from DR8 does not
  differ significantly from DR7.} (see Fig. \ref{fig-spec-werr}).  The
spectrum from INT (observed on 16th Dec. 1987) has been derived
directly from the plot given in \citet[][their
  Fig. 2]{1991-jackson-quasarprop}, thus it is suitable only for a
qualitative analysis.  The spectrum from SDSS (observed on 5th
Jan. 2006) has been automatically flux-- and wavelength--calibrated by
the SDSS pipeline \citep{2002-sdss-edr}, de--reddened using
\citet{1989-cardelli-extinction} with E(B--V)=0.0375
\citep{1998-Schlegel-galaxy-irmap}, transformed to rest frame of
reference using our redshift estimate ($z$=0.70747, \S
\ref{sec-spec-result}) and rebinned by a factor of 2
(Fig. \ref{fig-spec-werr}, see Fig. \ref{fig_fitsloan} for a closer
view of the H$\beta$ and [OIII] region).

\begin{figure*}
\includegraphics[width=12.8cm]{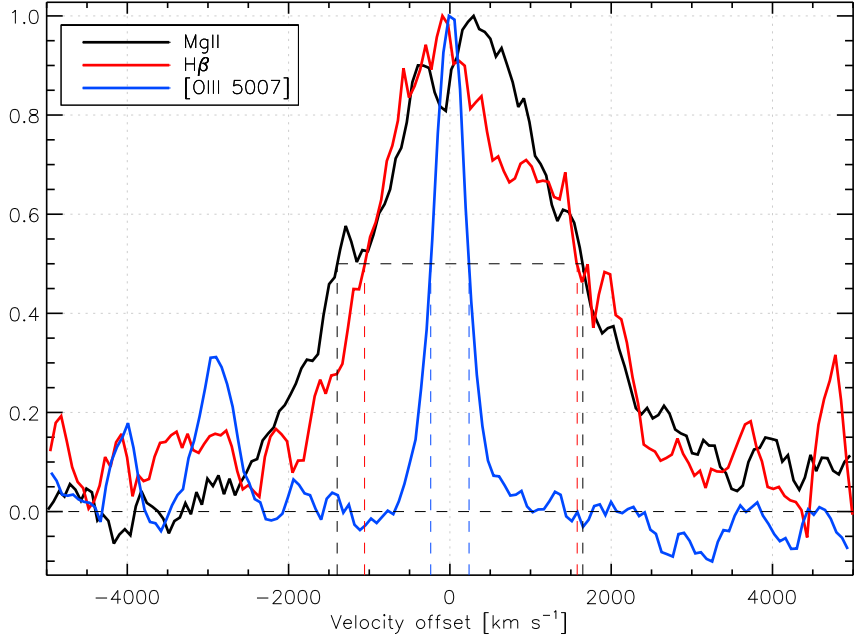}
\caption{Profiles of MgII (2800\AA), H$\beta$ (4863\AA) and [OIII]
  (5007\AA) emission lines in velocity offset space.  Spectrum has
  been continuum--subtracted, the continuum level has been estimated
  by eye and all profiles have been normalized to have maximum of 1.
  Finally, profiles have been smoothed by a $\sim$4 \AA \ boxcar
  average to provide a clearer view.  Dashed lines allow to roughly
  measure full width at half maximum (FWHM) of each line.}
\label{fig-fwhm}
\end{figure*}

Fig. \ref{fig-fwhm} shows the profiles of the main emission lines,
namely MgII (2800\AA), H$\beta$ (4863\AA) and [OIII] (5007\AA), in
velocity space.  The spectrum has been continuum--subtracted and all
profiles have been normalized to have maximum of 1.  Finally all
profiles have been smoothed by a $\sim$4 \AA\ boxcar average to provide
a clearer view.  The dashed lines allow to approximately measure the
full width at half maximum (FWHM) of each line.

To provide a better estimate of the FWHM of the H$\beta$ and [OIII]
line profiles we also performed a fit of the spectrum using five
components model (Model 1): a power--law to account for the continuum
emission, four Gaussian--shaped emission lines to account for the
H$\beta$ (both narrow and broad component) and the [OIII] 4959\AA,
[OIII] 5007\AA\ lines.  FWHM and velocity offset of the narrow lines
are forced to be the same.  Results are shown in
Fig. \ref{fig_fitsloan} (upper panel).  To account for residuals in
the 4830 -- 4900\AA\ range (i.e. the asymmetry in the broad H$\beta$
line) we ran another fit by adding another emission line component
named H$\beta^*$ (Model 2, Fig. \ref{fig_fitsloan} lower panel).

\begin{figure*}
\includegraphics[width=14cm]{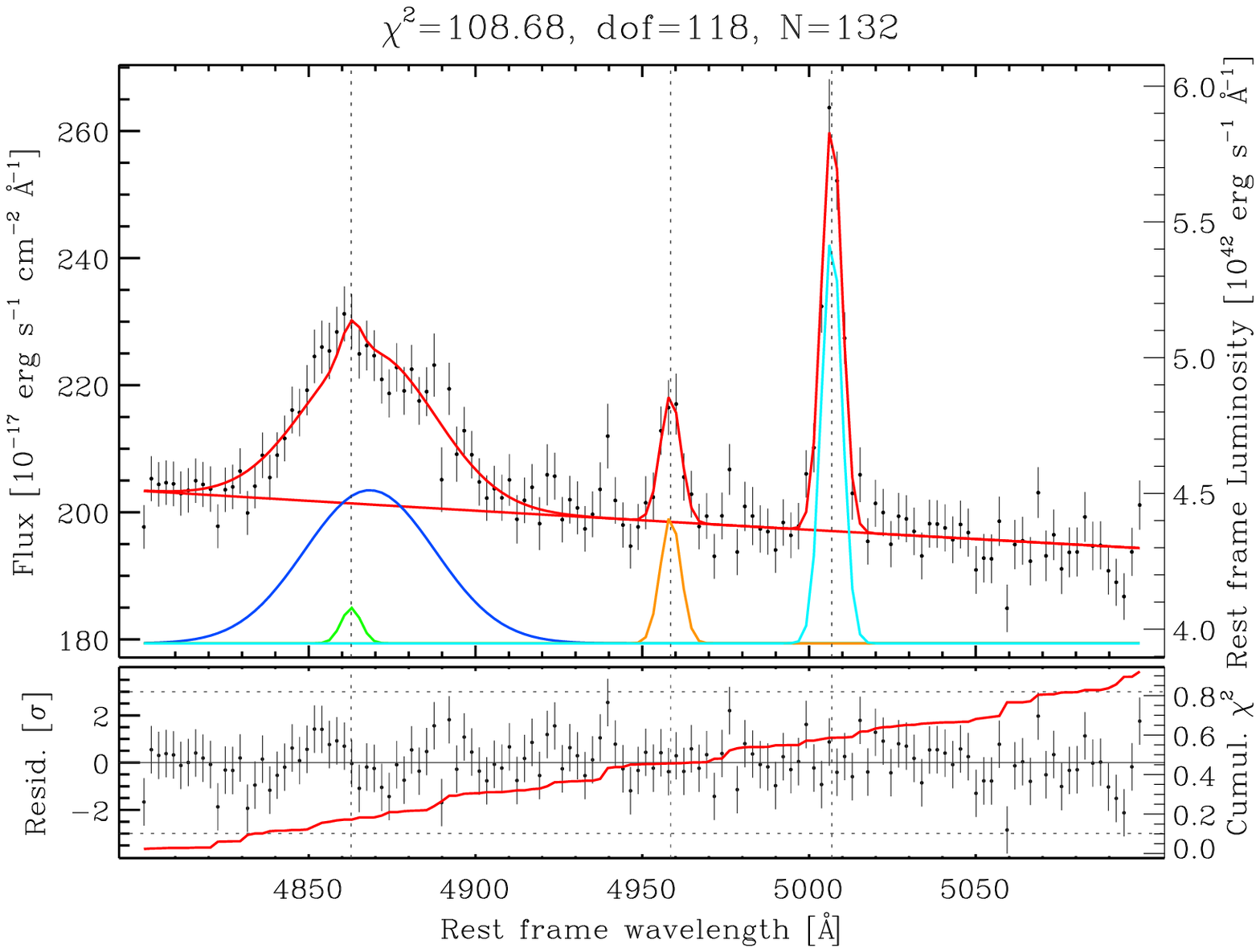}\\
\includegraphics[width=14cm]{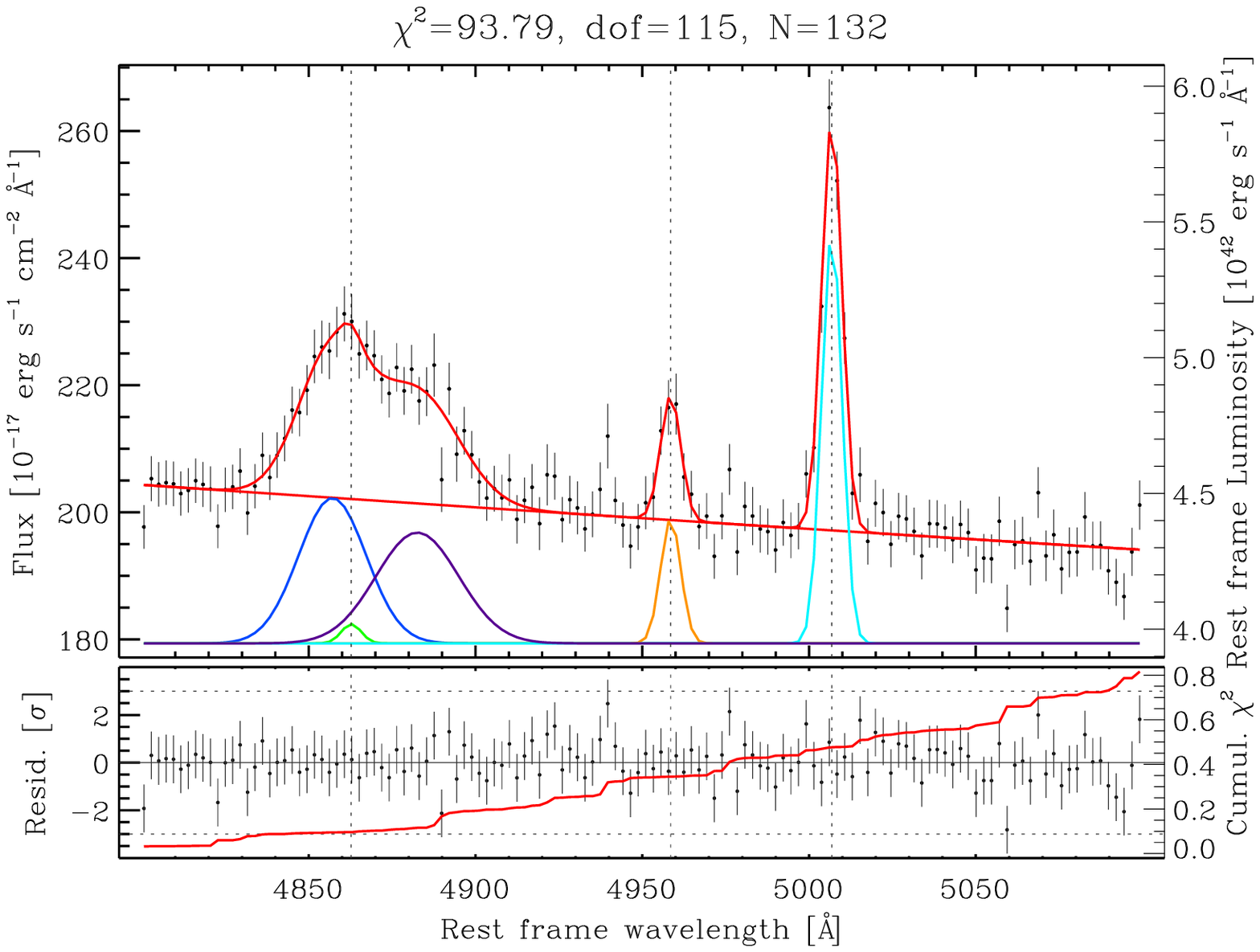}
\caption{Spectrum of \srcname\ and fitted model in the range
  4810--5100 \AA.  All quantities are quoted in rest frame.  Upper
  panel: spectrum (black points, 1$\sigma$ error bars) with fitted
  model (red points) and model components: broad lines (red solid
  line), narrow lines (blue solid lines) and power--law continuum
  (green solid line).  The broad and narrow lines components are
  shifted upward for clarity.  The five black dotted lines are the
  rest line centers of (from left to right) H$\gamma$, H$\beta$,
  ``line 2'' (see text), [OIII] 4959 and [OIII] 5007 respectively.
  Red and blue dotted lines are the rest line centers of all broad and
  narrow considered lines.  Lower panel: residual in units of one
  standard deviation.  Red solid line is the cumulative reduced
  chi--squared (values on right axis).}
\label{fig_fitsloan}
\end{figure*}

\subsection{Results}
\label{sec-spec-result}

A qualitative comparison of the spectra by INT and SDSS taken $\sim$9
years apart (Tab. \ref{tab-dateobs}) shows that the continuum has
changed both in intensity (by a factor $\sim$5) and in slope.  The
intensity and shape of the Balmer line seem to be the same in both
spectra.  In \S \ref{sub-sed-data} we also analyze the broad--band SED
(lower panel of Fig. \ref{fig-sed}) of the source, considering also
the photometric data from SDSS, taken $\sim$1 year before the SDSS
spectroscopic observations.  Again, we see that a change in intensity
of the optical emission of a factor $\sim$10 occurred in just 1 year.
Also, a change in slope occurred, as indicated by the photometric data
in the 5 available bands.  Thus, we conclude that \srcname\ has been
observed in at least three emission states at optical wavelengths,
which differs in intensity: from the dimmer one (SDSS/photometric),
through an intermediate one (INT), to the brighter one
(SDSS/spectroscopic).  The spectral slope is correlated to the change
of state.  As discussed in \S \ref{sec-sed-modeling-res}, the change
of state can be simply explained as a change of magnetic field. A
change in accretion rate is not needed, therefore we do not expect any
change in line luminosity and profile.  We will refer to these optical
emission states as the disk dominated one, the intermediate state and
the jet dominated state respectively, and discuss their properties in
\S \ref{sec-sed-modeling-res}.

A quantitative spectral analysis can only be performed on the SDSS
spectrum.  Fig. \ref{fig-fwhm} clearly shows that our value of
$z$=0.70747 locates the peak of the [OIII] line exactly at 0 in
velocity space, while the MgII peak is redshifted by $\sim$300 km
s$^{-1}$ and the H$\beta$ peak is blueshifted by $\sim$100 km
s$^{-1}$. Also, the H$\beta$ profile shows a pronounced asymmetry on
the red wing.  Rough estimates of the FWHMs are: 3050 km s$^{-1}$ for
the MgII line, 2600 km s$^{-1}$ for the H$\beta$ line and $\sim$480 km
s$^{-1}$ for the [OIII] line.

To provide better estimates of luminosities, widths and offsets of
emission lines, and eventually highlight some peculiarities in the
spectrum, we performed a fit with the models described above.  The
spectrum in the 4800 -- 5100\AA\ region, as well as the fitting model
and the single components, are shown in Fig. \ref{fig_fitsloan} (Model
1 in upper panel, Model 2 in lower panel).  Results are shown in
Tab. \ref{tab-data}.

{\bf Model 1:} The resulting FWHM for the H$\beta$ line is $(2830 \pm
220)$ km s$^{-1}$, while the FWHM of the narrow lines is $(430 \pm
24)$ km s$^{-1}$, grossly consistent with the rough estimates obtained
with Fig. \ref{fig-fwhm}.  The model fits reasonably well the
spectrum, and the $\chi^2$ value is rather small.  Also, the
luminosity of the continuum at 5100\AA\ and the spectral slope, as
well as the sum of luminosities from broad and narrow H$\beta$ lines
and [OIII] luminosities, are consistent with the fit performed in S10
(Tab. \ref{tab-data}), thus our model does not suffer from lack of
iron line modeling (which is present in their fit).  The value of FWHM
and velocity offset for both the broad and narrow components of
H$\beta$ are however, quite different.  The strong discrepancies in
FWHM between our results and those from S10 are probably due to the
fact that we had the opportunity to carefully check the results of the
fit, while in S10 an automatic fitting pipeline had to be used to
handle the many spectra contained in the catalog.

{\bf Model 2:} The residuals of the fit in Model 1
(Fig. \ref{fig_fitsloan}, upper panel) look random everywhere, except
in two regions: 4840--4900\AA\ and 4930--4945\AA, in which the
residuals are quite ``coherent'', suggesting the presence of further
components.  Although the model is far from being rejected
($\chi^2_{\rm red} < 1$) these residuals are due to the asymmetry in
the broad H$\beta$ profile discussed above, and to an additional
emission line.  Thus, we ran again the fit with a new broad
H$\beta$--like component, named H$\beta^*$ in Tab. \ref{tab-data}
(actually we are not interested in the additional line at 4940\AA).
Results are shown in Fig. \ref{fig_fitsloan} (lower panel) and
Tab. \ref{tab-data}.  The extra component accounts well for the
asymmetry of the H$\beta$ profile, and the residuals in the region now
look random.  The same consideration as for Model 1 applies for the
luminosity and slope of the continuum, and the luminosity of the
H$\beta$ and [OIII] lines.  Clearly, the putative broad H$\beta$ line
is now much narrower ($\sim$1500 km s$^{-1}$).  Furthermore, the new
H$\beta^*$ component has the same FWHM and velocity offset as the
broad H$\beta$ values given in S10.  This provides a possible
explanation for the discrepancies between our results and those from
S10.

\begin{table*}
 \centering
   \caption{Quantities derived from fit parameters and corresponding
     quantities in the \citet{2010-shen-catdr7} catalog.}
   \label{tab-data}
   \begin{tabular}{lr@{ $\pm$ }lr@{ $\pm$ }lr@{ $\pm$ }lc}
     \hline\hline
     \multicolumn{1}{c}{Parameter}                 &
     \multicolumn{2}{c}{Model 1}                   &
     \multicolumn{2}{c}{Model 2}                   &
     \multicolumn{2}{c}{S10}                       &
     \multicolumn{1}{c}{Units}                     \\
 Lum. (H$\beta_{\rm broad}$)             &        26.3 & 2.2        &       12.9 & 6.3        &    23 & 13             &   [10$^{42}$ erg s$^{-1}$]  \\
 Lum. (H$\beta^*$)                       &         \nan &           &       12.1 & 5.8        &     \nan &             &   [10$^{42}$ erg s$^{-1}$]  \\
 Lum. (H$\beta_{\rm narrow}$)            &        0.93 & 0.58       &       0.50 & 0.92       &    8 & 42              &   [10$^{42}$ erg s$^{-1}$]  \\
 Lum. ([OIII], 5007)                     &        11.07 & 0.59      &       11.01 & 0.58      &    12.2 & 1.9          &   [10$^{42}$ erg s$^{-1}$]  \\\hline
 FWHM (H$\beta_{\rm broad}$)             &        2830 & 220        &       1470 & 320        &    1870 & 600          &   [km s$^{-1}$]             \\
 FWHM (H$\beta^*$)                       &         \nan &           &       1790 & 600        &     \nan &             &   [km s$^{-1}$]             \\
 FWHM (H$\beta_{\rm narrow}$)            &        431 & 24          &       428 & 24          &    1200 & 400          &   [km s$^{-1}$]             \\\hline
 V$_{\rm off}$ (H$\beta_{\rm broad}$)    &        344 & 75          &       --(350 & 270)     &    --(1250 & 630)      &   [km s$^{-1}$]             \\
 V$_{\rm off}$ (H$\beta^*$)              &         \nan &           &       1220 & 400        &     \nan &             &   [km s$^{-1}$]             \\
 V$_{\rm off}$ (H$\beta_{\rm narrow}$)   &        --(0 & 10)        &       1 & 10            &    37 & 200            &   [km s$^{-1}$]             \\\hline
 $\lambda$L$_{\lambda}$ (5100\AA)        &        220.88 & 0.80     &       220.59 & 0.78     &    220.73 & 0.74       &   [10$^{44}$ erg s$^{-1}$]  \\
 Continuum index                         &        --(0.75 & 0.13)   &       --(0.85 & 0.11)   &    --(0.709 & 0.018)   &                             \\
     \hline\hline
   \end{tabular}
\end{table*}

In summary, our Model 2 seems to be the best description of the
spectrum of \srcname\ in the 4810 -- 5100 \AA\ range.

\subsection{The H$\beta$ line profile}
\label{sec-hbeta-profile}
The FWHM value of the H$\beta$ line profile reported in literature is
65 \AA\ \citep{1991-jackson-quasarprop}. As shown in
Tab. \ref{tab-data2} the FWHM found with both Model 1 and 2 of our fit
is significantly lower (46 \AA\ and 24 \AA\ respectively). Since the
H$\beta$ profiles observed with INT and SDSS
(Fig. \ref{fig-spec-werr}) overlap perfectly, we can exclude a
variation in the line profile, and conclude that previous value was
likely overestimated. Moreover, the value reported in S10 is perfectly
compatible with our estimate in Model 1 (48 \AA, Tab.\ref{tab-data2}).

Fig. \ref{fig_fitsloan} show that the broad H$\beta$ emission line can
hardly be modeled with a single Gaussian profile, nor with any
symmetric line profile such as the often quoted logarithmic profile
\citep{1975-Blumenthal-LogatihmicProfile}, since the red wing is
highly asymmetric.  Iron emission lines in the range 4840--4900
\AA\ are usually much weaker than the Balmer lines
\citep{1979-Capriotti-asymmetricLineProfilesInSy1}, thus such profile
is hardly the result of a blending of different lines. Iron lines may
be present but their intensity is negligible, indeed our fitting
procedure provide the same results for line luminosities and width of
narrow lines as reported in S10 (\S \ref{sec-spec-result}). Asymmetric
H$\beta$ line profiles are not new in Type 1 AGN spectra: both red and
blue asymmetries has been observed in all kind of AGNs
\citep[e.g.][]{1979-Capriotti-asymmetricLineProfilesInSy1,
  1980-Capriotti-DynamicsOfBLR, 1987-Peterson-DoubleBLRinNGC5548,
  1990-Stirpe-AtlasOfHa-Hb-Profiles, 1996-Romano-ha-profile-wings,
  2001-Veron-AtlasOfNLS1}, as well as profile variability
\citep{1977-Osterbrock-EmissionSpectraOfarakelian,
  1987-Peterson-VariabilityInNGC5548,
  1988-Stirpe-VariabilityInNGC5548-3783}. Several kinematic models
have been proposed to explain such features in the broad Balmer
spectra of AGN, including radial inflow or outflow, different
geometries and multiple components BLR
\citep[e.g.][]{1987-Peterson-VariabilityInNGC5548,
  2004-Popovic-ContributionOfDiskToBLR, 1996-Romano-ha-profile-wings,
  2009-Zhu-ILRInAGN}, but no one has yet reached a general consensus.
This hints for the presence of non virialized components in the BLR
region.  Thus, measures of FWHM performed on such profiles may not be
directly related to the underlying black hole mass.  This is the
reason why we used the FWHM of just the main H$\beta$ line to estimate
the virial black hole mass, neglecting the additional H$\beta^*$
component which account for the asymmetry.  Doing so, the FWHM of
H$\beta$ turns out to be about half the FWHM of MgII
(Fig. \ref{fig-fwhm}). Such differences are common in sources with
small values of FWHM(H$\beta$): in the S10 catalog, the MgII line is
wider than H$\beta$ for $\sim$84\% of the sources with
FWHM(H$\beta$)$<2000$, and only 22\% of the sources with
FWHM(H$\beta$)$>4000$.

Of course the reasoning can be inverted: the ``real'' H$\beta$ line
profile may be much wider, but some intervening gas could absorb
radiation only on the blue side.  Although we have no evidence to
exclude this hypothesis, the range of values for the black hole mass
found in \S \ref{sec-spec-derived} marginally supports the hypothesis
that the real virialized line profile is the one modeled with the main
H$\beta$ component.  This conclusion may have important consequences
on the classification of \srcname.  Should the additional H$\beta^*$
component vanish in future observations, the FWHM of the remaining,
virialized, H$\beta$ component would be $\sim$1500 km s$^{-1}$, and
the source would become a powerful
$\gamma$-RL-NLS1. \citep{2009-Foschini-discovery_pmnj0948,
  2009-Abdo-rlnls1_newclass, 2011-calderone-variab}.  Such variations
in line profiles have already been observed, on timescales of tens of
years in NGC 5548 \citep{1987-Peterson-VariabilityInNGC5548}.

\subsection{Spectroscopically derived quantities}
\label{sec-spec-derived}

Spectroscopic measurements, such as those derived in \S
\ref{sec-spec-result}, are used to infer some physical properties of
the source, such as the total accretion luminosity $L_{\rm bol}$ and
the virial black hole mass.  The former is obtained by scaling the
luminosity of one emission line (e.g. H$\beta$) to obtain the total
BLR luminosity, $L_{\rm BLR}$, through a template quasar spectrum
\citep{1991-francis-composite}.  $L_{\rm bol}$ can then be computed by
dividing $L_{\rm BLR}$ by the covering factor, assumed to be 0.1.
Black hole masses are usually estimated assuming virialized motion of
BLR clouds.  The BLR radius has been measured by means of the
reverberation mapping technique \citep[RM, ][]{1993-peterson-rm} for a
few tens of objects, and a correlation between the continuum optical
luminosity and the BLR size, has been calibrated \citep[$R$--$L$
  relation, ][]{2005-kaspi-relation, 2006-vp-masscalib,
  2009-bentz-calib-bhmass}, so that we can infer $R_{\rm BLR}$ size
without the (very time--consuming) RM observational campaigns.  Thus,
it is in principle possible to estimate black hole masses using
single--epoch optical spectroscopy (such as the SDSS spectrum of
\srcname\ used in \S \ref{sec-optspectro}) and a calibrated virial
mass scaling relationship (such as those reported in
\citet{2006-vp-masscalib}).  There are, however, several caveats, such
as the geometry and orientation of the BLR
\citep{2011-decarli-blrgeom}, the role of radiation pressure
\citep{2008-Marconi, 2009-Marconi-rpcorrectbhmass} and the way line
profiles should be analyzed to infer the orbital velocity of the
clouds.  Indeed, the H$\beta$ broad line is often asymmetric, thus the
broadening cannot be simply interpreted as being due to orbital bulk
motion.  In these cases is not clear what measure should be taken for
the line width.  Note also that RM studies have been carried out
mainly on radio--quiet sources, i.e. sources in which synchrotron
emission from a jet at optical/IR wavelengths is negligible with
respect to the emission from the so called Big Blue Bump (BBB).  Thus,
the $R$--$L$ relations are reliable only when the BBB is clearly
visible \citep{2004-Wu-R-L-RelationOnRadioLoudQSO}.  Contribution from
host starlight is also supposed to be an issue
\citep[e.g.][]{2009-bentz-calib-bhmass}, although only in low
luminosity AGN \citep[log($\lambda L_\lambda$,5100\AA)$<$44.5,
][]{2010-shen-catdr7}, therefore not in the case of \srcname.
Finally, the 1$\sigma$ statistical dispersion of virial mass scaling
relationships against RM--based masses is estimated to be $\sim$0.5
dex. The RM masses are in turn scattered by the same factor around the
$M$--$\sigma$ relation \citep{2004-Onken-virialVsMsigma}.  The
absolute accuracy is therefore estimated to be of the order of 0.7 dex
\citep{2006-vp-masscalib}.

We will consider virial mass estimates computed using both Model 1 and
2.  Also, the SDSS spectrum of \srcname\ has been taken when the
source was in the jet dominated state, therefore we cannot use the
continuum luminosity estimated from spectral analysis to infer the BLR
size.  Instead, we will use optical data of the disk dominated state,
as observed photometrically by SDSS.

\subsubsection{Results}
\label{sec-spec-derived-res}

To compute $L_{\rm BLR}$, and the disk bolometric luminosity $L_{\rm
  bol}$, we use the composite quasar spectrum given in
\cite{1991-francis-composite}, including the H$\alpha$ contribution as
in \citet{1997-Celotti-LblrVsJetPower}.  The broad H$\beta$ line
contributes to the entire $L_{\rm BLR}$ with a weight 22/555 (the
Ly$\alpha$ contributes for 100/555).  Using the luminosity of the
H$\beta$ complex found with Model 2 fitting (broad H$\beta$ line +
H$\beta^*$ component, log($L$(H$\beta$)/erg s$^{-1}$) = 43.4), and
setting $L_{\rm bol}= 10 L_{\rm BLR}$, we estimate log($L_{\rm
  bol}$/erg s$^{-1}$) = 45.8.  This value agrees with the one derived
in \S \ref{sec-sed-modeling} through SED modeling: log($L_{\rm
  bol}$/erg s$^{-1}$) = 46.  Instead, S10 reports log($L_{\rm
  bol}$/erg s$^{-1}$) = 47, because S10 used an observation made when
the source was in the jet dominated state.

\begin{table*}
 \begin{minipage}{140mm}
   \begin{center}
     \caption{Black hole mass estimates using different methods and
       calibrations.  Continuum luminosity at 3000\AA\ and
       5100\AA\ (rest frame) used in virial estimates are log($\nu
       L_\nu$/erg s$^{-1}$) = 45.7 and 45.3 for MgII and H$\beta$
       respectively (derived from SDSS photometry).  Col. [1]: method
       used; Col. [2]: reference for the FWHM of the emission line;
       Col. [3]: FWHM of the broad emission line; Col. [4]: black hole
       mass estimate; Col. [5]: Eddington ratio, assuming a bolometric
       luminosity log($L_{\rm bol}$/erg s$^{-1}$) = 45.95 (see \S
       \ref{sec-sed-modeling} and Tab. \ref{para}); Col. [6]:
       calibration references: (a) \citet{2010-shen-catdr7}, (b)
       \citet{2009-vestergaard-calib-bhmass}, (c)
       \citet{2006-vp-masscalib}, (d) \citet{2009-bentz-calib-bhmass},
       (e) \citet{2002-tremaine-velocity-o3}.}
     \label{tab-data2}
     \begin{tabular}{llcccc}
       \hline\hline
       \multicolumn{1}{c}{Method}               &
       \multicolumn{1}{c}{FWHM ref.}            & 
       \multicolumn{1}{c}{FWHM [km s$^{-1}$, (\AA)]}   & 
       \multicolumn{1}{c}{log M/M$_{\sun}$}     & 
       \multicolumn{1}{c}{Edd. ratio}           &
       \multicolumn{1}{c}{Calib. ref.}          \\
       \multicolumn{1}{c}{[1]} &
       \multicolumn{1}{c}{[2]} &
       \multicolumn{1}{c}{[3]} & 
       \multicolumn{1}{c}{[4]} & 
       \multicolumn{1}{c}{[5]} &
       \multicolumn{1}{c}{[6]} \\
       \hline
       \multirow{2}{*}{Virial, MgII}                    & S10, broad comp.        &   3980     ( 65   )    &   9.0    &   0.1      & (a) \\
                                                        & S10, whole profile      &   3390     ( 55   )    &   8.8    &   0.1      & (b) \\\hline
       \multirow{2}{*}{Virial, H$\beta$, single comp.}  & S10                     &   2970     ( 48   )    &   8.6    &   0.2      & (c) \\
                                                        & Model 1                 &   2830     ( 46   )    &   8.5    &   0.3      & (d) \\\hline
       \multirow{1}{*}{Virial, H$\beta$, H$\beta^*$}    & Model 2                 &   1470     ( 24   )    &   7.9    &   1.1      & (d) \\
       M-$\sigma_*$ ([OIII])                            & Model 2                 &   430      ( 7    )    &   8.0    &   0.9      & (e) \\
       SED modeling                                     & \nan                    &    \nan                &   8.2    &   0.6      &     \\
     \hline\hline
   \end{tabular}
   \end{center}
 \end{minipage}
\end{table*}

Tab. \ref{tab-data2} reports the black hole mass estimates for
\srcname\ obtained using different methods and calibrations. For all
estimates, except the last two, we used the virial mass scaling
relations.  The first two estimates are computed using the FWHM
estimates (of the broad component and the whole profile respectively)
of the MgII emission line provided by S10, the continuum luminosity
derived from SDSS photometry (when the source was in the disk
dominated state) of log($\nu L_\nu$/erg s$^{-1}$) = 45.7 and the same
calibrations used in S10: \citet{2010-shen-catdr7} and
\citet{2009-vestergaard-calib-bhmass} respectively.  The mass
estimates are very similar, so we will refer to them as the MgII
virial mass estimates (log $M/M_{\sun}\sim 9$).

The next two estimates are computed using the FWHM of the H$\beta$
emission line fitted with a single Gaussian.  Values for FWHM are
provided by S10 and by our Model 1 respectively, the continuum
luminosity is derived from SDSS photometry (disk dominated state):
log($\nu L_\nu$/erg s$^{-1}$) = 45.3, and the calibration are from
\citet[][same as S10]{2006-vp-masscalib} and
\citet{2009-bentz-calib-bhmass}, respectively.  This pair of mass
estimates are also very similar, so we will refer to them as the
single H$\beta$ virial mass estimates (log $M/M_{\sun}\sim 8.5$).

Next estimate is computed using the FWHM of the H$\beta$ emission line
from Model 2, neglecting the additional H$\beta^*$ component, and the
calibration by \citet{2009-bentz-calib-bhmass}.

Finally, the last two rows of Tab. \ref{tab-data2} report mass
estimates obtained without using the virial method.  For the first one
we used the FWHM of the [OIII] line (provided by Model 2 fitting) as a
proxy for the stellar velocity dispersion, and the calibration of the
$M-\sigma_*$ relation given in \citet[][their
  Eq. 20]{2002-tremaine-velocity-o3}.  Updated calibration given in
\citet{2011-Graham-Msigma-calib} produces the same result.  In the
last one we simply report the mass estimate obtained in \S
\ref{sec-sed-modeling} through SED modeling.  Notice that the last
three methods yield very similar results (log $M/M_{\sun}\sim 8$),
even if they have been derived using different and independent
methods.

\subsubsection{Discussion}
\label{sec-disc-bhmass}
Our black hole mass estimates are considerably lower than those found
in the literature (\S \ref{sec-intro-b2}). This is likely a consequence of
the different values of the H$\beta$ line profile's width, as
discussed in \ref{sec-hbeta-profile}. Also, the mass given in S10 is
probably overestimated since the continuum luminosity has been
measured while the source was in the jet dominated state. Therefore,
we will consider only our mass estimates given in
Tab. \ref{tab-data2}.

In particular, the estimate obtained with Model 2 (neglecting the
H$\beta^*$ component), and the $M$-$\sigma$ relation
(log($M/M_{\sun}$) $\sim$8) are quite independent, thus they
provide greater confidence about the goodness of the mass estimate.
Modeling the disk dominated state with a standard
\citep[i.e. ][]{1973-ssad} disk (\S \ref{sec-sed-modeling-res}) yields
estimates of the black hole mass between $(1.5-5)\times 10^8 M_\odot$.
Smaller masses are excluded because the disk would become
super--Eddington, larger masses are excluded because, beside providing
a worse fit to the near--IR and optical data of the
``disk--dominated'' state, they require a smaller ionizing luminosity,
that becomes incompatible with the luminosity of the broad line
region.  We therefore conclude that the black hole mass of
\srcname\ is in the range log($M/M_{\sun}$) = 8--8.5.

\section{Modeling the SED}
\label{sec-sed-modeling}

To model the SED we use the one--zone leptonic model described in
detail in \citet{2009-Ghisellini-Canonical_blazar}, whose main
characteristics are briefly described in the Appendix.  The SED of the
source is shown in Fig. \ref{fig-sed}.  The lower panel shows a zoom
of the optical, X--ray and $\gamma$--ray data part of the SED,
together with the models used to interpret the two states of the
source.  As discussed above, SDSS observed \srcname\ in two occasions,
approximately one year apart.  In the first observation the SDSS made
the photometry in its usual optical filters (black squares in
Fig. \ref{fig-sed}), while the spectrum, one year later, revealed a
very high state, dominated by the beamed synchrotron emission (orange
lines), in agreement with the GALEX observations (orange open
circles). 

For the low optical state we have simultaneous UVOT and XRT {\it
  Swift} observations: the UVOT point (blue filled circle) agrees with
the photometric SDSS data points (black squares) and with the 2MASS
data (red triangles).  We have then modeled these data together with
the XRT X--ray data, assuming also that the average {\it Fermi}/LAT
flux and spectrum is a good representation of the high energy emission
of this state.

On the other hand we have no information about the level of the
$\gamma$--ray flux for the high optical state of the source.  We
therefore assume that it corresponds also to a high state of the
X--ray emission, that can be represented by the maximum X--ray flux
observed (violet diamond point), and to the same $\gamma$--ray flux.

The {\it Chandra} and {\it Fermi} data shown as green/black ties
are simultaneous (even if the {\it Fermi} spectrum is the average of
34 months of observations), but we do not have any information about
the optical--UV state.  During the 34 months of {\it Fermi}
observations, the $\gamma$--ray flux was rather stable, with no sign
of significant variability.
The one--zone leptonic model (in which optical and $\gamma$--ray
radiation are produced by the same population of electrons) is
perfectly suitable to describe this SED.  Detection of optical
variability without corresponding changes in simultaneous
$\gamma$--ray observations would be, however, difficult to accomodate
in the framework of the one--zone leptonic model.

In Tab. \ref{para} we list the relevant parameters used to interpret
the two states of the source, namely the disk and the jet dominated
one.  Tab. \ref{powers} reports the jet powers corresponding to the
two states, calculated as explained in the Appendix.

\subsection{Guidelines for the choice of the parameters}

{\it Accretion disk} --- The luminosity of the accretion disk can be
estimated through the luminosity of the broad emission lines $L_{\rm
  BLR}$, assuming a given value for the covering factor.  As discussed
in detail in \S \ref{sec-spec-derived}, we can estimate $L_{\rm BLR}
\sim 6.3\times 10^{44}$ erg s$^{-1}$. Assuming a covering factor of
$\sim$0.1, the disk luminosity is of the order of $10^{46}$ erg
s$^{-1}$.  Given the uncertainties about $L_{\rm BLR}$, derived using
only the $H\beta$ line, the covering factor and the mass of the black
hole (controlling the $L_{\rm d}/ L_{\rm ion}$ ratio), this estimate
necessarily has an uncertainty of at least a factor 2.

\vskip 0.3 cm
\noindent
{\it Location of the emitting region} --- The SED in the low state
shows that the high energy hump dominates the bolometric output, as in
other powerful Flat Spectrum Radio Quasars (FSRQs).  In these objects
it is likely that the emitting region is located inside the BLR, since
in these cases the seed photons produced by the BLR, whose flux is
seen enhanced in the comoving frame of the emitting region, dominate
the radiative cooling.  We then assume $R_{\rm diss}<R_{\rm BLR}$,
where $R_{\rm diss}$ is the distance of the emitting region from the
black hole, and $R_{\rm BLR}\sim 10^{17} L_{\rm d, 45}^{1/2}$ cm is
the size of the BLR.

\vskip 0.3 cm
\noindent
{\it Magnetic field} --- The value of the magnetic field is chosen in
order to reproduce the ratio between the synchrotron and the inverse
Compton flux.

\vskip 0.3 cm
\noindent
{\it Bulk Lorentz factor} --- This has been chosen in order to
reproduce the superluminal apparent velocity ($\beta_{\rm app} \sim
12$, \citealt{2004-Kellermann-SubMasImagingOfQSO}).  This also fixes
the viewing angle, that must be of the order of $1/\Gamma$.  Note that
the value of $\Gamma$ determines the value of the radiation energy
density of the external seed photons ($\propto \Gamma^2$) and hence
the value of the magnetic field required to have the observed
synchrotron to inverse Compton luminosity ratio.

\vskip 0.3 cm
\noindent
{\it Injected power} --- The power is injected throughout the source
in the form of relativistic electrons. Through the continuity equation
we calculate the particle distribution as a result of injection,
cooling and possible pair production.  The total injected power is
such that the radiation produced by these particles agrees with the
observed data.  The injected distribution is assumed to be a smoothly
broken power law.  The resulting distribution, modified by cooling,
must agree with the observed slopes.

\subsection{Results}
\label{sec-sed-modeling-res}
The most notable feature of \srcname\ is the large variability in the
optical, from a disk to a jet dominated emission.  This is interpreted
as a change in the magnetic field, from $\sim 5$ to 20 G, accompanied
by an increase of the total injected power (by a factor 2), and a
decrease in the maximum particles energy and a flattening of the high
energy slope of the particle distribution.  The location of the
emitting region is somewhat less in the high optical state.  We
maintained the same bulk Lorentz factor.  The power of the jet of
\srcname\ carried by the magnetic field is larger in the high optical
state (factor 10), while the differences in the other forms of power
are minor.  The jet power ($P_{\rm p} + P_{\rm e}$) is dominated by
the bulk motion of protons.  An electron to (cold) proton ratio of
$\sim 1$ results in $P_{\rm p} \sim (2-5) \times 10^{47}$ erg.  The
jet power decreases by a factor of 20 if there are 20
electron--positron pairs per proton.  Such a lepton to proton ratio is
about the maximum allowed to make the jet not to recoil under the
Compton drag effect \citep{2010-Ghisellini-jetminpower}.  A value
$P_{\rm p}\sim 2\times 10^{46}$ erg s$^{-1}$ makes the jet power
similar to the accretion disk luminosity (which is slightly larger
than $10^{46}$ erg s$^{-1}$).  The minimum value of jet power is the
one needed to produce the observed radiation, that is of the order of
$P_{\rm r} \sim L_{\rm obs}/\Gamma^2$= (3--7)$\times 10^{44}$ erg
s$^{-1}$.  Even this absolute lower limit on the jet power contrasts
with the absence of a strong extended emission (see \S 3).  In fact,
following \citet{1999-Willott-NLR-radio-correlation}, we can estimate
the jet power $Q$ from the extended luminosity at 151 MHz.
Extrapolating from 5 GHz, assuming $F_{\nu} \propto \nu^{-1}$, we
derive $ Q\sim 5 \times 10^{43}$ erg s$^{-1}$, a value that is even
smaller than $P_{\rm r}$.

These values of the physical parameters are similar to those found by
fitting {\it Fermi} blazars \citep[][]{2010-Ghisellini-blazarprop},
with the black hole mass being in the low end of the distribution.  We
have used a black hole mass $M=1.5\times 10^8 M_\odot$.  We have also
fitted the spectrum with larger black hole masses, and obtain equally
good fits with $M=3\times 10^8 M_\odot$ (and $L_{\rm d}/L_{\rm
  Edd}=0.34$) and $M=5\times 10^8 M_\odot$ (and $L_{\rm d}/L_{\rm
  Edd}=0.17$).  Both these alternative solutions are acceptable since
they provide a sufficient $L_{\rm ion}$ to photoionize the BLR.  If
$M=10^9 M_\odot$, instead, the computed disk emission shifts to
smaller frequencies (i.e. the disk becomes colder) and then we are
obliged to decrease its overall luminosity.  As a consequence, we
cannot reproduce the {\it Swift}/UVOT $u$ point, and the ionizing
luminosity becomes too small to efficiently photoionize the BLR.

The picture that emerges is that of a typical {\it Fermi} blazar, with
a flat radio spectrum produced by a powerful jet, and black hole mass
significantly smaller than $10^9 M_\odot$.  The source periodically
cycles through different emission states, characterized by different
amount of power emitted by the jet.  Although the SED parameters are
only rough estimates of physical quantities, the possibility to model
two different states of the same source provides at least a clue that
the parameters kept fixed in both models (black hole mass, disk
luminosity, BLR radius, covering factor) are reliable.

{\small
\begin{table*} 
\centering
\begin{tabular}{lllllllllllllll}
\hline
\hline
State  &$R_{\rm diss}$ &$M$ &$L_{\rm d}$ &$R_{\rm BLR}$ &$P^\prime_{\rm i}$  
&$B$ &$\Gamma$ &$\theta_{\rm v}$ 
    &$\gamma_{\rm b}$ &$\gamma_{\rm max}$ &$s_1$  &$s_2$ &$\gamma_{\rm cool}$ \\
~[1]     &[2] &[3] &[4] &[5] &[6] &[7] &[8] &[9] &[10] &[11] &[12]  &[13] &[14] \\
\hline   
jet dominated    &45 (1000)  &1.5e8  &11.3 (0.58)  &335 &0.014 &21   &13   &3  &100 &2.5e3 &2    &2.5  &7   \\
disk dominated   &59 (1300)  &1.5e8  &11.3 (0.58)  &335 &7e--3 &5.2  &13   &3  &100 &5e3   &2    &2.9  &21  \\\hline
PMN J0948+0022   & 2 (1600)  &1.5e8  & 9 (0.4)     &300 &0.024 &3.4  &10   &6  &800 &1.6e3 &1    &2.2  &    \\
3C 273           &120 (500)  &8e8    &48 (0.4)     &693 &0.015 &11.6 &12.9 &3  & 40 &2e4   &1    &3.4  &    \\
\hline
\hline 
\end{tabular}
\vskip 0.4 true cm
\caption{Input parameters used to model the SED of \srcname\ (first
  two lines), PMN J0948+0022 and 3C 273 \citep{2010-Ghisellini-blazarprop}.
Col. [1]: State of the source;
Col. [2]: distance of the blob from the black hole in units of $10^{15}$ cm and in units of the \sc\ radius;
Col. [3]: black hole mass in solar masses;
Col. [4]: Disk luminosity in units of $10^{45}$ erg s$^{-1}$ and in Eddington units;
Col. [5]: distance of the BLR in units of $10^{15}$ cm;
Col. [6]: power injected in the blob calculated in the comoving frame, in units of $10^{45}$ erg s$^{-1}$; 
Col. [7]: magnetic field in Gauss;
Col. [8]: bulk Lorentz;
Col. [9]: viewing angle in degrees;
Col. [10], [11] and: minimum and maximum random Lorentz factors of the injected electrons;
Col. [12] and [13]: slopes of the injected electron distribution [$Q(\gamma)$] below 
and above $\gamma_{\rm b}$;
Col. [14] minimum
random Lorentz factor of the electrons cooling in one light crossing time $R/c$.
The disk has an X--ray corona of luminosity $L_X=0.3 L_{\rm d}$.
The spectral shape of the corona is assumed to be $\propto \nu^{-1} \exp(-h\nu/150~{\rm keV})$.
}
\label{para}
\end{table*}

\begin{table} 
\centering
\begin{tabular}{llllll}
\hline
\hline
State   &$\log P_{\rm r}$ &$\log P_{\rm B}$ &$\log P_{\rm e}$ &$\log P_{\rm p}$ &$\log (P_{\rm p}/20)$ \\
\hline  
jet domin.      &44.81 &45.75  &44.92  &47.71  &46.41 \\
disk domin.     &44.56 &44.77  &44.71  &47.41  &46.11 \\
\hline
0948+0022       &45.30 &44.35  &44.71  &46.68  &45.38 \\
3C 273          &45.05 &46.09  &44.90  &47.48  &46.18 \\
\hline
\hline 
\end{tabular}
\vskip 0.4 true cm
\caption{Jet power in the form of radiation, Poynting flux, bulk
  motion of electrons and protons (assuming one cold proton per
  emitting electron and 20 pairs per proton) for of \srcname\ (first
  two lines), PMN J0948+0022 and 3C 273
  \citep{2010-Ghisellini-blazarprop}.  }
\label{powers}
\end{table}
}

\section{Comparison with 3C 273 and PMN J0948+0022}
\label{sec-sed-compare}
\begin{figure}
\includegraphics[width=8cm]{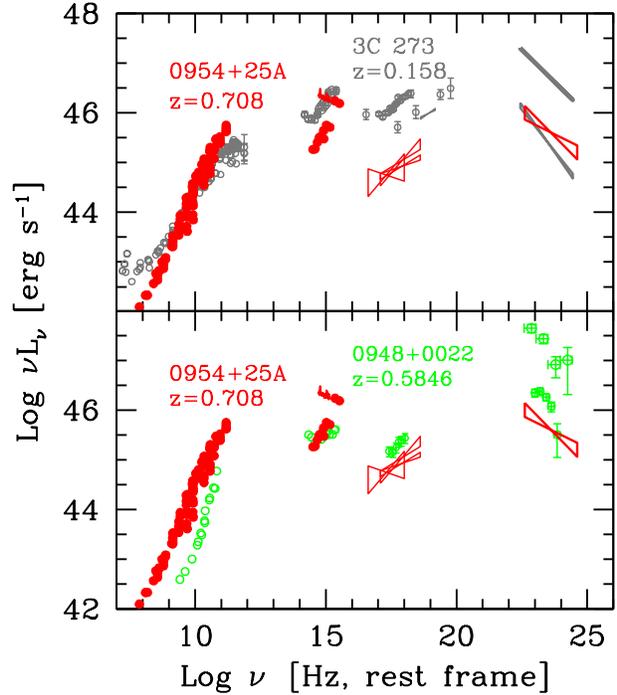}
\caption{Comparison of the SED of \srcname\ with the SED of 3C 273
  (top panel) and PMN J0948+0022 (bottom).}
\label{fig-cmp273-0948}
\end{figure}

In Fig. \ref{fig-cmp273-0948} we compare the broad band SED of
\srcname\ to that of 3C 273 (upper panel) and PMN J0948+0022 (lower
panel). Both sources have been analyzed in
\citet{2010-Ghisellini-blazarprop} and model parameters are reported
in Tab. \ref{para} and \ref{powers}.  The radio spectrum of the
prototypical blazar 3C 273 shows a flat spectrum comparable in
luminosity with \srcname, although slightly steeper.  At the lowest
observed frequencies the two SED are markedly different: whereas the
74--365 MHz radio spectral index of 3C 273 become steep ($\alpha \sim
-0.5$), the spectral index of \srcname\ remains flat ($\alpha \sim
-0.2$). As discussed in \S \ref{sec-radioprop-0954}, this behaviour
can be explained with a different amount of relativistic beaming which
boosts the jet emission in \srcname. At shorter wavelengths 3C 273 is
$\sim$1 order of magnitude more powerful than \srcname.  If emission
in both sources occur at approximately the same Eddington ratio
(0.4--0.6, Tab. \ref{para}) as suggested by SED modeling, the
luminosity ratio at optical/UV wavelengths ($\sim$ 3) provides an
estimate for the mass ratio.  Since log($M/M_{\sun}$) of 3C 273 is
8.9, for \srcname\ we expect a mass of $\sim$8.4.  This is yet another
indication that the black hole mass of \srcname\ should be
significantly smaller than log($M/M_{\sun})=9$.

Also, the SED of PMN J0948+0022 is very similar to that of
\srcname\ (in the disk dominated state), the only difference being at
$\gamma$--rays (where PMN J0948+0022 showed an exceptional flaring
episode \citep{2010-Foschini-outburst}).  Parameters from the SED
modeling and jet powers are quite similar (Tab. \ref{para} and
\ref{powers}).  This similarities support our speculation on the
possible classification of \srcname\ as a $\gamma$--NLS1.

However, this similarity cannot be pushed too far, since
\srcname\ lacks the strong iron emission that usually characterizes
NLS1.  Also, X--ray photon index of \srcname is $\Gamma=1.74$ (\S
\ref{sub-sed-data}), while that of NLS1 is typically $\Gamma>2.5$
\citep{1996-boller-nls1-softx}.  However, consider that also the other
NLS1 detected by {\it Fermi} have a flat X--ray spectrum
\citep{2009-Abdo-mw_monitor_pmnj0948, 2009-Abdo-rlnls1_newclass}
likely because the flat X--ray jet emission dominates.

\subsection{Typical {\it Fermi} blazar or $\gamma$-NLS1?}
\label{sec-nls1}

As discussed in \S \ref{sec-sed-modeling} and \S
\ref{sec-sed-compare}, \srcname\ is a typical {\it Fermi}
blazar.  However, the likely small black hole mass of \srcname\ and the
SED resemblance with prototypical $\gamma$--NLS1 PMN J0948+0022
suggest some similarities between the two sources.  \srcname\ does not
meet the commonly adopted classification criterion for NLS1, namely a
FWHM of broad H$\beta<$2000 km s$^{-1}$
\citep{1985-Osterbrock-spectra_of_nls1}.\footnote{The further
  criterion on flux ratio [OIII]/H$\beta<$3 is almost always
  automatically satisfied when a broad H$\beta$ component is detected
  \citep[e.g. ][]{2006-Zhou-comprehensivestudy}.} However, the
physical nature of such empirical threshold has been often questioned
\citep[e.g. ][]{1989-Goodrich-spectropolarimetry-NLS1,
  2001-Veron-AtlasOfNLS1} since all observational properties show a
continuous transition at FWHM(H$\beta) \sim$2000 km s$^{-1}$,
i.e. properties of NLS1 and BLS1 sources are smoothly joined.

Alternative criteria to distinguish among BLS1 and NLS1 sources have
been put forward. \citet{2000-Sulentic-EV1} proposed to increase the
dividing threshold to 4000 km s$^{-1}$, to select radio--quiet sources
with low mass and high accretion rate. \citet{2001-Veron-AtlasOfNLS1}
suggested to consider the strength of FeII emission relative to
H$\beta$ as a possible tracer of the Eddington ratio.
\citet{2007-Netzer-evolOfAccretionMetal_inAGN} proposed to classify as
narrow--line AGN (NLAGN) all sources exceeding an Eddington ratio of
0.25 (regardless of black hole mass).

A small black hole mass is likely a characterizing property of NLS1
sources \citep{2004-grupe-mass-sigma-NLS1}.  We would like to stress
that if asymmetry in broad Balmer line profiles are due to
non-virialized components in the BLR, then only the virialized one
should be considered to estimate the black hole mass, and thus the
NLS1 classification.  In this case \srcname\ would be classified as a
powerful $\gamma$--NLS1, just like PMN J0948+0022.

Since the overall properties of \srcname\ are similar to both classes
of {\it Fermi} blazars and $\gamma$--NLS1, we conclude that it is one
of those objects which smoothly joins the population of narrow--line
and broad--line AGN.

\section{Summary and conclusions}
In this paper we carried out an extensive analysis on the {\it
  Fermi} blazar \srcname\ using archival data. We
  highlighted several 
peculiarities in the aim at providing a coherent picture of its physical
properties. The main conclusions are as follows:
\begin{itemize}

\item (\S \ref{sec-intro-b2}, \S \ref{sec-data-fermi}) the source
  \srcname\ is a strong $\gamma$--ray emitter, detected both in the
  1yr and 2yr {\it Fermi}/LAT point source catalogs \citep{2010-1FGL,
    2011-lat2fgl}. We find no significant variability of the flux nor
  of the spectral index in the 0.1--100 GeV band in the first 34
  months of data from {\it Fermi}/LAT. 

\item (\S \ref{sec-radioprop-0954}) its radio spectrum is ``flat''
  down to very low frequencies \citep[74 MHz,][]{2007-VLA-survey},
  i.e. the optically thick synchrotron emission from the jet is
  dominating over the optically thin, isotropic emission from extended
  region. This observation indicates that the viewing angle must be
  small (3--6 degrees).  The corresponding de--projected size of the
  jet is in the range 0.5--1 Mpc, the size of a giant radio lobe. The
  isotropic and the beamed components should have equal luminosities
  at $\sim$10 MHz.
  
\item (\S \ref{sec-radioprop-0954}, Fig. \ref{fig-radiospindex}) the
  ``flatness'' of the radio spectrum at low frequencies observed in
  \srcname\ is typical of {\it Fermi} detected blazars (2LAC sample),
  i.e. well aligned blazars.  Radio selected blazars show steeper
  spectra on average (e.g. the CRATES sample).

\item (\S \ref{sec-spec-result}, Fig. \ref{fig-sed} and
  \ref{fig-spec-werr}) \srcname\ has been observed in at least three
  different optical states: from a disk--dominated state, to a jet
  dominated one in which the synchrotron emission overwhelms the
  emission from the disk, via an intermediate state.  The two extreme
  states differ by nearly one order of magnitude in luminosity.
  Despite the change in continuum luminosity, the broad hydrogen
  Balmer lines maintain the same luminosity and line profile.

\item (\S \ref{sec-spec-result}, Fig. \ref{fig_fitsloan},
  Tab. \ref{tab-data}) the previously known estimate for the FWHM of
  the broad H$\beta$ line profile (65 \AA\ rest frame, corresponding
  to $\sim$4000 km s$^{-1}$) from \citet{1991-jackson-quasarprop} is
  probably overestimated. Also, the value of 1870 km s$^{-1}$ reported
  in \citet{2010-shen-catdr7} is probably due to a poor model fit. A
  more likely value is 46 \AA, corresponding to $\sim$2800 km
  s$^{-1}$.  The broad H$\beta$ line shows a pronounced asymmetry on
  the red wing, which require a second Gaussian component (H$\beta^*$)
  of similar FWHM, to be modeled.  The velocity offset of the
  H$\beta^*$ is $\sim$1200 km s$^{-1}$.

\item (\S \ref{sec-spec-result}, Fig. \ref{fig-fwhm}) [OIII]
  5007\AA\ and MgII lines are quite symmetric. The FWHM of the MgII
  line is $\sim$ 3050 km s$^{-1}$, and the peak is redshifted by
  $\sim$300 km s$^{-1}$ with respect to the redshift identified by the
  [OIII] line.

\item (\S \ref {sec-spec-derived}, \S \ref{sec-disc-bhmass}) the most
  likely value of the black hole mass is $M\sim (1-3)\times 10^8
  M_\odot$, as suggested by several independent methods. In
  particular, for the mass scaling relations we considered only the
  symmetric component of the H$\beta$ profile, and the continuum
  luminosity in the disk dominated state.

\item (\S \ref{sec-sed-modeling}, Fig. \ref{fig-sed}, Tab. \ref{para},
  \S \ref{sec-sed-compare}) modeling the broad band spectral energy
  distribution, we derive physical parameters quite typical of other
  {\it Fermi} blazars.

\item (\S \ref{sec-sed-compare}) We suggest to classify \srcname\ as a
  transition object between the class of FSRQ and $\gamma$-NLS1, since
  it shows characteristic features of both classes (namely, the {\it
    blazar} appearance and the similarity with the SED of PMN
  J0948+0022).
\end{itemize}

\appendix
\section{Appendix: Some details on the modeling}

We use the model described in detail in
\citet{2009-Ghisellini-Canonical_blazar}.  The emitting region is assumed
spherical, at a distance $R_{\rm diss}$ from the black hole, of size $R=\psi
R_{\rm diss}$ (with $\psi=0.1$), and moving with a
bulk Lorentz factor $\Gamma$.  The bolometric luminosity of the accretion disk
is $L_{\rm d}$.

The energy particle distribution $N(\gamma)$ [cm$^{-3}$] is calculated
solving the continuity equation where particle injection, radiative
cooling and pair production (via the $\gamma$--$\gamma \to e^\pm$
process), are taken into account.  The created pairs contribute to the
emission.  The injection function $Q(\gamma)$ [cm$^{-3}$ s$^{-1}$] is
assumed to be a smoothly joining broken power--law, with a slope
$Q(\gamma)\propto \gamma^{-{s_1}}$ and $\gamma^{-{s_2}}$ below and
above a break energy $\gamma_{\rm b}$:
\begin{equation}
Q(\gamma)  \, = \, Q_0\, { (\gamma/\gamma_{\rm b})^{-s_1} \over 1+
(\gamma/\gamma_{\rm b})^{-s_1+s_2} }
\label{qgamma}
\end{equation}
The total power injected into the source in the form of relativistic
electrons is $P^\prime_{\rm i}=m_{\rm e}c^2 V\int Q(\gamma)\gamma
d\gamma$, where $V=(4\pi/3)R^3$ is the volume of the emitting region.

The injection process lasts for a light crossing time $R/c$, and we
calculate $N(\gamma)$ at this time.  This assumption comes from the
fact that even if injection lasted longer, adiabatic losses caused by
the expansion of the source (which is traveling while emitting) and
the corresponding decrease of the magnetic field would make the
observed flux to decrease.  Therefore the calculated spectra
correspond to the maximum of a flaring episode.

Above and below the accretion disk, in its inner parts, there is an
X--ray emitting corona of luminosity $L_{\rm X}$ (it is fixed at a
level of 30\% of $L_{\rm d}$).  Its spectrum is a power law of energy
index $\alpha_X=1$ ending with a exponential cut at $E_{\rm c}=$150
keV.  The specific energy density (i.e. as a function of frequency) of
the disk and the corona are calculated in the comoving frame of the
emitting blob, and used to properly calculate the resulting External
inverse Compton spectrum.  The BLR is assumed to be a thin spherical
shell, of radius $R_{\rm BLR}=10^{17} L_{\rm d, 45}^{1/2}$ cm.
We consider also the presence of a IR torus, at larger distances.  The
internally produced synchrotron emission is used to calculate the
synchrotron self Compton (SSC) flux.  Table \ref{para} lists the
adopted parameters.

Table \ref{powers} lists the power carried by the jet in the form of
radiation ($P_{\rm r}$), magnetic field ($P_{\rm B}$), emitting
electrons ($P_{\rm e}$, no cold electron component is assumed) and
cold protons ($P_{\rm p}$, assuming one proton per emitting electron).
All the powers are calculated as
\begin{equation}
P_i  \, =\, \pi R^2 \Gamma^2\beta c \, U^\prime_i
\end{equation}
where $U^\prime_i$ is the energy density of the $i$ component, as
measured in the comoving frame.

The power carried in the form of the produced radiation, $P_{\rm r}
=\pi R^2 \Gamma^2\beta c \, U^\prime_{\rm rad}$, can be re--written as
[using $U^\prime_{\rm rad}=L^\prime/(4\pi R^2 c)$]:
\begin{equation}
P_{\rm r}  \, =\,  L^\prime {\Gamma^2 \over 4} \, =\, L {\Gamma^2 \over 4 \delta^4}
\, \sim \, L {1 \over 4 \delta^2}
\end{equation} 
where $L$ is the total observed non--thermal luminosity ($L^\prime$ is
in the comoving frame) and $U^\prime_{\rm rad}$ is the radiation
energy density produced by the jet (i.e.  excluding the external
components).  The last equality assumes $\theta_{\rm v}\sim 1/\Gamma$.

When calculating $P_{\rm e}$ (the jet power in bulk motion of emitting
electrons) we include their average energy, i.e.  $U^\prime_{\rm e}=
n_{\rm e} \langle\gamma\rangle m_{\rm e} c^2$.

\bibliography{agn}
\bibliographystyle{mn2e}

\bsp

\label{lastpage}
\end{document}